\documentclass[pra,amssymb,amsmath,onecolumn,a4paper]{revtex4}

\usepackage{bm}
\usepackage{graphicx}

\newcommand{\beq}{\begin{equation}}
\newcommand{\eeq}{\end{equation}}
\newcommand{\beqa}{\begin{eqnarray}}
\newcommand{\eeqa}{\end{eqnarray}}

\newcommand{\Hh}{l}
\newcommand{\Ll}{m}
\newcommand{\Mm}{m}
\newcommand{\Nn}{n}
\newcommand{\Qq}{q}
\newcommand{\pe}{p}

\newcommand{\kt}{k_\perp}

\begin{document}

\title{Mode density of orbital angular momentum modes in a cylindrical cavity and in free space}

\author{Mauritz Andersson, Eilert Berglind and Gunnar Bj\"{o}rk}

\affiliation {Department of Applied Physics, Royal Institute of Technology (KTH)\\
AlbaNova University Center, SE-106~91 Stockholm, Sweden.\\}

\date{\today}

\begin{abstract}
In this short note we derive an approximate expression for the mode density of modes carrying orbital angular momentum (OAM) in a cylindrical cavity which is large compared to the inverse wavevector in any direction. We argue that in the large cavity limit the modes asymptotically converges to the corresponding OAM modes in free space. We compare the result to Weyl's law. It is found, as expected, that the mode density does not depend on whether or not the modes carry OAM.
\end{abstract}

\maketitle

\section{Introduction}
Recently, the interest in using the angular momentum degree of freedom in both classical and quantum optics has increased. The attraction for  both communities is the anticipated possibility to utilize angular momentum as a new degree of freedom. Claims have been made that using the orbital angular momentum degree of freedom will increase the communication capacity per unit frequency and unit volume in radio links \cite{Tamburini}. We cite from the abstract of \cite{Tamburini}: ``This novel radio technique allows the implementation of, in principle, an infinite number of channels in a given, fixed bandwidth, even without using polarization, multiport or dense coding techniques.'' This claim has generated considerable excitement in the physics, electromagnetics, and microwave engineering communities \cite{Nature,Spectrum,Telegraph} and therefore it merits a critical examination.

 Weyl's law states that any set of orthogonal modes have the same density of states provided the ``cavity'' or volume one confine the modes into is large enough. In fact, the requirement is that for any spatial co-ordinate $\hat{q}$, the cavity dimension $q$ in that direction times the wave vector in the same direction $k_q$ should fulfil $q k_q \gg 1$. ``Free space'' certainly fulfills the condition as one may make a fictious cavity arbitrarily large and let it have an arbitrary shape.

To exemplify Weyl's law we derive the mode density for two sets of modes. In Sec. \ref{Sec: DOS cubic} we derive the mode density for a cubic cavity. Although the final result is very well known from the literature we do this by splitting the density of states into a transverse and a longitudinal factor. Such a division simplifies the corresponding derivation in Sec. \ref{Sec: DOS cylinder}, where we derive the mode density of angular momentum modes in a cylindrical cavity and show that the mode density shows no increase. The latter modes are essentially equal the OAM Bessel modes in free space, so the claim of an communication capacity increase in free space is unfounded.

\section{Density of states in a cubic cavity}
\label{Sec: DOS cubic}
To quickly derive a frame of reference without consideration of angular momentum, we start by calculating the density of modes in an empty, ideal, metallic cavity formed like a cube with the side $L$. The modes in such a cavity are TE and TM modes satisfying the dispersion relation
\beq
\left ( \frac{\omega}{c}\right )^2 = \left ( \frac{\pi \Mm}{L}\right )^2 + \left ( \frac{\pi \Nn}{L}\right )^2 + \left ( \frac{\pi \Qq}{L}\right )^2,
\eeq
where $\omega$ is the angular frequency, $c$ is the speed of light, and $\Mm, \Nn$, and $\Qq$ are 0, 1, 2, \ldots but two of them cannot be zero at the same time. Hence, along all axes in cartesian $\overline{k}$-space the mode spacing is $\pi/L$.

Now let us divide the $\overline{k}$-vector into a transverse and a longitudinal direction. We will arbitrarily take $k_z$ as the longitudinal wavevector component and the component lying in the $xy$-plane to be the transverse component $\kt$. The number of TE-modes $N_\perp$ having a transverse wavevector smaller or equal to $\kt$ is found by dividing the circular k-space area $\pi \kt^2$ with the transverse $k$-vector ``unit cell'' $(\pi/L)^2$, that is:
\beq
N_\perp = \frac{\pi \kt^2 L^2}{\pi^2}.
\eeq
Hence,
\beq
\frac{d N_\perp}{d \kt}=  \frac{2  L^2 \kt}{ \pi}.
\eeq
The mode density in the longitudinal direction is $d N_z/d k_z = L/\pi$. To get the total number $N$ (i.e., both TE and TM) of modes with a wavevector shorter or equal to $k$ we integrate the mode density over all such wave vectors. We keep in mind that $k^2 = \kt^2 + k_z^2$.
\beqa
N & = & 2 \cdot \frac{1}{8} \int_{-k}^k dk_z \int_0^{\sqrt{k^2 - k_z^2}} d \kt \frac{L}{\pi} \frac{2  L^2 \kt}{ \pi}  \nonumber \\
&=&  \frac{L^3}{4 \pi^2} \int_{-k}^k dk_z \left ( k^2 - k_z^2 \right )= \frac{L^3 k^3}{3 \pi^2},
\eeqa
where the pre-factor 2 in front of the first integral accounts for the inclusion of TM modes, and the subsequent factor 1/8 is due to that only modes in the first octant of $\overline{k}$-space should be counted since $\Mm, \Nn$, and $\Qq$ are positive integers. Thus, we get the mode density
\beq
\frac{d N}{d k} = \frac{V k^2}{\pi^2},
\label{Eq: Exact derivation}
\eeq
where $V=L^3$ is the cavity volume. This density can trivially be recast as
$$
\frac{d N}{d \nu} = \frac{8 \pi V \nu^2}{c^3} \ \ \  \mbox{\textrm{or}} \ \ \  \frac{d N}{d \omega} = \frac{V \omega^2}{\pi^2 c^3},
$$
where $\nu$ denotes frequency, $\omega$ denotes angular frequency, and $c$ is the speed of light in vacuum. The only approximation in (\ref{Eq: Exact derivation}) is due to the discrete nature of $\Mm, \Nn$, and $\Qq$ when the cavity side length $L$ is on the order of a wavelength (so that $V k^3$ is on the order of unity). When $V k^3 \gg 1$, (\ref{Eq: Exact derivation}) is exact.

\section{Orbital angular momentum modes in a cavity with circular cross-section}
\label{Sec: DOS cylinder}
The modes inside a hollow, circular, metallic waveguide can carry angular momentum in the form of orbital angular momentum and spin. In many ways they are very similar to the modes of a Bessel beam (that also can carry OAM) and when the cavity dimensions are large compared to $c/\nu$ the solutions are essentially identical to Bessel beams in free space \cite{Berglind}. Therefore, under this condition the mode density is essentially equal for the free-space modes and the cavity modes. However, we find that the cavity mode density is easier to derive which is why we look at this case.

In a circular waveguide with radius $a$ the TE modes are given by the expressions (in cylindrical coordinates) \cite{Collins}:
\beqa
H_z & = & -\frac{i Ak_0^2}{\mu_0 \omega} J_\Hh\left ( \frac{\pe'_{\Hh \Ll} \rho}{a}\right ) e^{(i\Hh \varphi)}  e^{(i[\omega t - \beta_{\Hh \Ll} z])}, \label{Eq: H_z circ}\\
E_\rho & = &  - \frac{ i A k_0^2 \Hh}{k_{\Hh \Ll}^2 \rho} J_\Hh\left ( \frac{\pe'_{\Hh \Ll} \rho}{a}\right )e^{(i\Hh \varphi)}
e^{(i[\omega t - \beta_{\Hh \Ll} z])}, \label{Eq: erho}\\
E_\varphi & = &    \frac{A k_0^2 }{k_{\Hh \Ll}} J'_\Hh\left ( \frac{\pe'_{\Hh \Ll} \rho}{a}\right )e^{(i\Hh \varphi)} e^{(i[\omega t - \beta_{\Hh \Ll} z])}, \label{Eq: Ephi}
\eeqa
where $A$ is an amplitude (with SI unit Volt), $\mu_0$ denotes the permeability of vacuum, $J'_\Hh(x)$ is the first derivative of the $\Hh$:th Bessel function of the first kind,  $\pe'_{\Hh \Ll}$ gives the $\Ll$:th solution to $J'_\Hh(x)=0$, $k_{\Hh \Ll}=\pe'_{\Hh \Ll}/a$ and $k_0= \omega/c$. One can show that these eigenmodes carry the angular momentum $\Hh \hbar$ per photon, where $\Hh$ is an integer (but can take on both positive and negative values) and furthermore, when $\pe'_{\Hh \Ll}-\Hh^2 \gg 1$ the angular momentum is primarily carried as orbital angular momentum \cite{Berglind}. It is important to remark that if one pair-wise superimposes such modes with equal amplitude and radial mode index but opposite chirality ($\Hh$ and $-\Hh$), then these new modes carry no OAM. The in-phase superposition will furthermore be orthogonal to the out-of-phase superposition so we will get an OAM-free, complete set of modes with pairwise correspondence to the OAM modes (except for the $\Hh=0$ mode that has no OAM anyhow).

If we make a cavity of such a waveguide of length $L$ by sealing both ends with a perfectly conducting plate, then in the longitudinal direction the cavity eigensolutions will be standing waves with a longitudinal mode density $L/\pi$.
To count the number of modes in the radial direction we will make some approximations of the Bessel functions of the first kind, since the functions and their zeros are not easily expressible analytically. First we note that $J_\Hh(x)$ and $J'_\Hh(x)$ are oscillating around zero with roughly the period $\pi$ for large values of $x$ (this is a slight underestimation of the period). This means that the spacing between consecutive $\pe'_{\Hh \Ll}$ (and also the corresponding zeros for $J_\Hh(x)$) for a given $\Hh$ must also be very nearly $\pi$. This gives us the mode spacing between consecutive transverse wavevectors, e.g., $(\pe'_{\Hh \Ll+1}-\pe'_{\Hh \Ll})/a=\pi/a$ which translates to a mode density of $a/\pi$.

\begin{figure}[ht]
\includegraphics[scale=0.9]{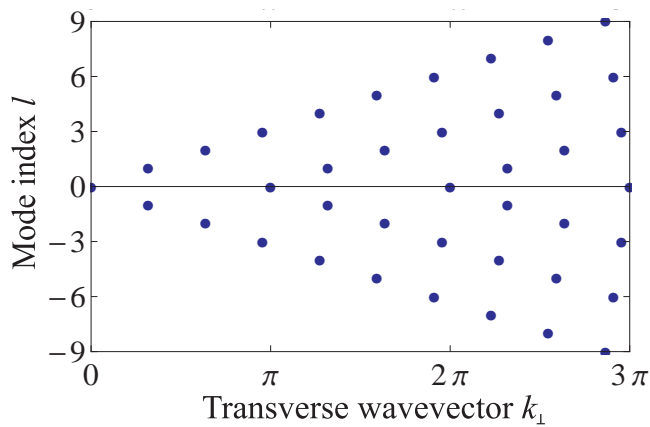}
\caption{The cylindrical cavity modes marked in a figure where the horizontal coordinate gives the transverse wavevector $\kt$ and the vertical coordinate gives the OAM mode index $\Hh$.}
\label{Fig: 1}
\end{figure}

Secondly, the first zero of $J_\Hh(x)$ and $J'_\Hh(x)$ occurs roughly when $x=\Hh$. For example, the first zero of $J_{2000}(x)$ occurs for $X \approx 2023,46$ and the first zero of $J_{2001}(x)$ occurs for $X \approx 2024,47$. Thus, with an increment of $1/a$ in wavevector space, one will exceed the wavevector cut-off for another series of mode solutions with even higher angular momentum. Summing the number of modes with a transverse wavevector no larger than $\kt$ one gets (see Fig. \ref{Fig: 1})
\beq
N_\perp = 2 \frac{\kt^2}{(\pi/a) (1/a)} = \frac{2 a^2 \kt^2}{\pi},
\eeq
where the pre-factor 2 again takes into account that there is also an equally large set of TM solutions \cite{Remark}. Hence, the transverse mode density is
\beq
\frac{d N_\perp}{d \kt} = \frac{4 a^2 \kt}{\pi}.
\eeq
The total number of modes are thus
\beq
N  =  \frac{4 a^2 L }{\pi^2} \int_0^k dk_z \int_0^{\sqrt{k^2 - k_z^2}} d \kt \kt  = \frac{4 a^2 L k^3}{3 \pi^2}.
\eeq
Taking the cavity volume $V=\pi a^2 L$ into account, the mode density approximated this way becomes
\beq
\frac{d N}{d k} = \frac{4 V k^2}{ \pi^3}=\frac{4}{\pi} \cdot \frac{V k^2}{\pi^2}.
\label{Eq: OAM mode density}
\eeq
We see that except for the pre-factor $4/\pi \approx 1.27$ we arrive at the same result as from the previous derivation. The discrepancy between (\ref{Eq: Exact derivation}) and (\ref{Eq: OAM mode density}) is not surprising, because we have used approximations to derive the mode spacing for these angular momentum modes whereas the derivation of (\ref{Eq: Exact derivation}) was exact (under the condition stated in Sec. \ref{Sec: DOS cubic}). That our estimate is an overestimate is expected, because the distance between successive zeros for $J'_\Hh(x)$ is in fact always larger than $\pi$, and as we have seen above, the first zero of the function occurs not when $x=\Hh$, but for a somewhat larger value. Therefore we have underestimated the mode spacing leading to an overestimation of the mode density. This is reflected by the numerical pre-factor $4/\pi$ that is an artefact due to our approximations.

\section{Conclusions}
We have derived an approximate expression for the mode density for a set of complete, orthogonal modes in a cavity with a circular cross section. These modes carry angular momentum and the mode index $\Hh$ directly gives the number of angular momentum quanta per photon the modes carry. As explained above, this mode set therefore very closely approximates free-space Bessel beams when the cavity dimensions are much larger than $c/\nu$ and thus should have the same mode density. We find that the mode density of these modes has the same scaling with frequency, angular frequency or wave vector as, for example, a plane wave set of modes would have. From this we conclude that the angular momentum carried by electro-magnetic waves does not increase the channel capacity compared to a mode set void of orbital angular momentum. (Such a mode set could easily be constructed from equally weighted sums and differences of angular momentum modes with indices $\Hh$ and $-\Hh$ \cite{Berglind}). Using one set of eigenmodes one codes the information in another degree of freedom than with the other set, but the number of power-orthogonal modes (or ``channels'') per unit frequency bandwidth remains the same.

\acknowledgments
GB thanks the Max-Planck-Institute f\"{u}r die Physik des Lichts for hosting him during the fall 2014. This work was economically supported by a Wenner-Gren Foundation stipend and by the Swedish Research Council through the Linn\ae us Excellence Center ADOPT through contract 349-2007-8664.


\begin{thebibliography}{1}
\bibitem{Remark} The TM modes can be derived from the dual set of equations to (\ref{Eq: H_z circ})-(\ref{Eq: Ephi}), i.e. with the $E$- and $H$-fields interchanged. However, the radial mode solutions in the TM case are given by the zeros of $J_\Hh(x)$ and not those of $J'_\Hh(x)$. However, using these TM modes and the same argumentation as for the TE modes also leads to (\ref{Eq: OAM mode density}).
\end{thebibliography}
\end{document}